\documentclass[a4paper,10pt]{article}

\usepackage{amsthm,amsmath,amssymb,amsfonts}

\def\s3{\sqrt{3}}

\newtheorem{The}{Theorem}

\DeclareMathOperator{\sech}{sech}

\begin{document}

\vspace{-2cm}
\title{Integrable systems of the intermediate long wave type in $2+1$ dimensions}

\author{B. Gormley$^1$, E.V. Ferapontov$^{1,2}$,  V.S. Novikov$^{1}$,
M.V. Pavlov$^{3}$}
     \date{}
     \maketitle
\begin{center}
$^1$Department of Mathematical Sciences, Loughborough University \\
Loughborough, Leicestershire LE11 3TU, United Kingdom \\
    \ \\
$^2$Institute of Mathematics, Ufa Federal Research Centre\\
Russian Academy of Sciences, 112 Chernyshevsky Street \\
Ufa 450008, Russian Federation\\
    \ \\
$^3$Lebedev Physical Institute \\
Russian Academy of Sciences,\\
Leninskij Prospekt 53, 119991 Moscow, Russia\\
\ \\
e-mails: \\[1ex]
\texttt{B.Gormley@lboro.ac.uk}\\
\texttt{E.V.Ferapontov@lboro.ac.uk}\\
\texttt{mpavlov@itp.ac.ru} \\
\texttt{V.Novikov@lboro.ac.uk}\\

\end{center}


\begin{abstract}
We classify  $2+1$ dimensional integrable systems with nonlocality of the intermediate long wave type. Links to the $2+1$ dimensional waterbag system 
are established.
Dimensional reductions of integrable systems constructed in this paper provide dispersive regularisations of  hydrodynamic equations governing  propagation of long nonlinear waves in a  shear flow with piecewise linear velocity profile (for special values of vorticities).
\end{abstract}


\bigskip


\section{Introduction }

The starting point of this paper is the $2+1$ dimensional integrable  equation
\begin{equation}\label{IL}
u_t=uu_y+\frac{\epsilon}{2}\frac{T+1}{T-1}u_{yy},
\end{equation}
which first appeared in \cite{Date} as a differential-difference  KP equation, see also \cite{Tamizhmani}.  Here  $u=u(x, y, t)$ and the symbol $T$ denotes  $\epsilon$-shift  in the $x$-variable: $Tu(x, y, t)=u(x+\epsilon, y, t)$.   
Introducing the nonlocal variable $w=\frac{\epsilon}{2}\frac{T+1}{T-1}u_{y}$, known as nonlocality of the intermediate long wave (ILW) type, we can represent equation (\ref{IL})  in  the form
\begin{equation}\label{IL1}
u_t=uu_y+w_y, \qquad \triangle w=\frac{T+1}{2}u_y,
\end{equation}
where $ \triangle=\frac{T-1}{\epsilon}$ is the discrete $x$-derivative.  In the dispersionless limit, $\epsilon \to 0$, we obtain the system $u_t=uu_y+w_y, \ w_x=u_y$ which is equivalent to a single PDE,
$u_t=uu_y+\partial_x^{-1}u_{yy}$, discussed in \cite{Zakharov1} as a counterpart of the dispersionless KP equation.

Generalising example (\ref{IL1}), paper \cite{FNR} gives a classification of $2+1$ dimensional integrable equations  of ILW  type, 
\begin{equation}
u_t=\varphi(u, w)u_x+\psi(u, w) u_y +\tau (u, w) w_x+ \eta (u, w)w_y+\epsilon(...)+\epsilon^2(...), \qquad \triangle w=\frac{T+1}{2}u_y.
\label{ILWtype1}
\end{equation}
Here  dots at $\epsilon$ and $\epsilon^2$ denote terms which are homogeneous differential polynomials of degree two and three in the $x-$ and $y-$derivatives of $u$ and $w$, respectively (the coefficients of these polynomials are allowed to be arbitrary functions of $u$ and $w$).   It turns out that $\epsilon$-terms, as well as all terms containing derivatives with respect to $x$ (in particular, coefficients $\varphi$ and $\tau$), must vanish identically, leading to the following classification result.

\begin{The} \label{th1} \cite{FNR} The following  examples constitute a complete list of integrable equations  (\ref{ILWtype1})  of ILW type:
\begin{eqnarray}
u_t&=&uu_y+w_y,  \label{ilw1}\\ 
u_t&=&(w+\alpha e^u)u_y+w_y,  \label{ilw2} \\
u_t&=&u^2u_y+(uw)_y+\frac{\epsilon^2}{12}u_{yyy},   \label{ilw3} \\
u_t&=&u^2u_y+(uw)_y+\frac{\epsilon^2}{12}\left(u_{yy}-\frac{3}{4}\frac{u_y^2}{u}\right)_y.  \label{ilw4}
\end{eqnarray}
In all cases the  nonlocality is the same: $\triangle w=\frac{T+1}{2}u_y$.
\end{The}

\noindent Equation (\ref{ilw3}) is a differential-difference analogue of the Veselov-Novikov equation discussed in \cite{Qian}. Equation (\ref{ilw4})  can be viewed as a differential-difference version of the modified Veselov-Novikov equation. Note that although equations 
(\ref{ilw1}) and (\ref{ilw2}) do not contain $\epsilon$-terms, they should still be considered as dispersive  due to the form of  the nonlocality.
 
The above classification utilised the perturbative approach of \cite{FerM, FMN} based on the requirement that all hydrodynamic reductions \cite{Fer4} of the dispersionless limit can be deformed into reductions of the full dispersive equation. This method provides an efficient approach to the classification of integrable systems in $2+1$ dimensions.

 Lax pairs, dispersionless limits and dispersionless Lax pairs of equations from Theorem 1 are  provided in the table below; note that  equations (\ref{ilw3}) and (\ref{ilw4}) have coinciding dispersionless limits/dispersionless Lax pairs. Here and in what follows, Lax pairs are obtained as quantisations of dispersionless Lax pairs   \cite{Zakharov}.
\begin{center}
\begin{tabular}{ | l | l | l | p{3cm} |} \hline
 $Eqn$ & $Lax~pair$ & $Dispersionless$ & $Dispersionless$ \\ 
 $$ &  & $limit$ & $Lax~pair$  \\ \hline 
 &  &  & \\
(\ref{ilw1}) & $T\psi=\epsilon \psi_y-u\psi$ & $u_t=u u_y+w_y$ & $e^{S_x}=S_y-u$ \\
 & $\epsilon \psi_t=\frac{\epsilon^2}{2}\psi_{yy}+(w-\frac{\epsilon}{2}u_y)\psi$ & $w_x=u_y$ & $S_t=\frac{1}{2}S_y^2+w$ \\ 
 &  &  & \\ \hline
 &  &  & \\
(\ref{ilw2}) & $T\psi=\epsilon e^{-u}\psi_y-\alpha \psi  $ & $u_t=(w+\alpha e^u)u_y+w_y$ & $e^{S_x}=e^{-u}S_y-\alpha$ \\
 & $\psi_t=\frac{\epsilon}{2}\psi_{yy}+(w-\frac{\epsilon}{2}u_y)\psi_y$ & $w_x=u_y$ & $S_t=\frac{1}{2}S_y^2+wS_y$ \\ 
 &  &  & \\ \hline
  &  &  & \\
(\ref{ilw3}) & $\epsilon(T-1)\psi_y=-2u(T+1)\psi$ & $u_t=u^2u_y+(uw)_y$ & $\frac{e^{S_x}-1}{e^{S_x}+1}S_y=-2u$ \\
 & $\psi_t=\frac{\epsilon^2}{12}\psi_{yyy}+(w-\frac{\epsilon}{2}u_y)\psi_y$ & $w_x=u_y$ & $S_t=\frac{1}{12}S_y^3+wS_y$ \\ 
 &  &  & \\ \hline
&  &  & \\
(\ref{ilw4}) & $\epsilon(T-1)\psi_y=\frac{\epsilon}{2}\frac{u_y}{u}(T-1)\psi-2u(T+1)\psi $ & $u_t=u^2u_y+(uw)_y$ & $\frac{e^{S_x}-1}{e^{S_x}+1}S_y=-2u $ \\
& $\psi_t=\frac{\epsilon^2}{12}\psi_{yyy}+(w-\frac{\epsilon}{2}u_y)\psi_y+\frac{1}{2}(w_y-\frac{\epsilon}{2}u_{yy})\psi$  & $w_x=u_y$ & $S_t=\frac{1}{12}S_y^3+wS_y$ \\ 
 &  $$ &  &  \\ \hline
\end{tabular}
\end{center}

\subsection{Summary of the main results}

As a generalisation of ansatz (\ref{ILWtype1}), in this paper we classify  integrable systems of the form
\begin{equation}\label{ILW}
\begin{array}{l}
\vspace{2mm}
u_t=\alpha(u,v,w)u_y+\beta(u,v,w)v_y+\gamma(u,v,w)w_y+\epsilon(\ldots),\\ 
\vspace{2mm}
v_t=\phi(u,v,w)u_y+\psi(u,v,w)v_y+\eta(u,v,w)w_y+\epsilon(\ldots),\\
\triangle w=\frac{T+1}{2}u_y.
\end{array}
\end{equation}
Here  dots at $\epsilon$  denote  homogeneous differential polynomials of degree two  in the $y$-derivatives of $u, v$ and $w$, whose coefficients are allowed to be arbitrary functions of $u, v, w$ (one can show that the coefficients at $x$-derivatives have to vanish identically). 
This classification problem has several important differences from that of Theorem \ref{th1}. First of all, no cases of Theorem \ref{th1} contain $\epsilon$-terms (thus, integrable systems of type (\ref{ILW}) constitute an essentially multi-component phenomenon). 
Secondly, there exist no integrable systems of type (\ref{ILW}) without $\epsilon$-terms, that is, no analogues of cases (\ref{ilw1}), (\ref{ilw2}) of Theorem \ref{th1}. In what follows, we will assume that the dispersion relation of the dispersionless limit of system (\ref{ILW}) is an irreducible cubic curve (see Section \ref{sec:proof}).

\begin{The}
\label{th2}
Modulo point transformations $v\to f(u,v)$, rescaling of $u, v$ and $w$, and Galilean transformations, there exist six integrable  systems of the form (\ref{ILW}):

\begin{eqnarray}
\label{ILW1}
&&
\begin{array}{l} 
u_t=(uv)_y-\epsilon u_{yy},
\\ v_t=uu_y+vv_y+4w_y+\epsilon v_{yy},
\end{array}\\ \nonumber \\ \nonumber \\
\label{ILW2}
&&\begin{array}{l} 
u_t=wu_y+v_y+\frac{\epsilon}{2} u_{yy},
\\ v_t=(vw+\frac{1}{2}v^2)_y-\frac{\epsilon}{2} (v_{yy}-(vu_y)_y),
\end{array}\\ \nonumber \\\nonumber \\
\label{ILW3}
&&\begin{array}{l} 
u_t=wu_y+v_y+\frac{\epsilon}{4} u_{yy},
\\ v_t=\frac{v^2}{\cosh^2 u}u_y+(w+2v\tanh u)v_y+vw_y-\frac{\epsilon}{4} \left(v_{yy}-\frac{1}{2}v\tanh u\ u_{yy}-\frac{v}{2\cosh^2u}u_y^2-\frac{1}{2}\tanh u\ u_yv_y\right),
\end{array}\\ \nonumber \\\nonumber \\
\label{ILW4}
&&\begin{array}{l} 
u_t=uu_y+vv_y+w_y+\frac{\epsilon}{4} v_{yy},
\\ v_t=(uv)_y-\frac{\epsilon}{4} u_{yy},
\end{array}\\ \nonumber \\\nonumber \\
\label{ILW5}
&&\begin{array}{l} 
u_t=uu_y+v_y+w_y+\frac{i}{4}\epsilon u_{yy},
\\ v_t=2vu_y+uv_y-\frac{i}{4}\epsilon (v_{yy}+w_{yy}-u_y^2),
\end{array}\\ \nonumber \\\nonumber \\
\label{ILW6}
&&\begin{array}{l} 
u_t=(ve^u+w)u_y+e^uv_y+w_y+\frac{\epsilon}{4} \left(\frac{v_y}{\sqrt{v^2+\alpha}}\right)_y,
\\ v_t=e^u(v^2+\alpha)u_y+(e^uv+w)v_y-\frac{\epsilon}{4} \left(\sqrt{v^2+\alpha}(u_{yy}+u_y^2)-\frac{v_y^2}{\sqrt{v^2+\alpha}}\right).
\end{array}\\ \nonumber 
\end{eqnarray}
In all cases the  nonlocality is the same: $\triangle w=\frac{T+1}{2}u_y$.
\end{The}

The proof of Theorem \ref{th2} is summarised in Section \ref{sec:proof}.

Dispersionless limits of some of the above systems 
can be obtained as special cases of the $2+1$ dimensional waterbag system, 
$$
u^{i}_{t}=u^iu^{i}_{y}+w_y,\qquad
w_{x}=\underset{k=1}{\overset{N}{\sum}}\frac{1}{\omega_{k}}u^{k}_{y}, 
$$
$i=1, \dots, N$, which is discussed in Section \ref{sec:wb}. We show that the waterbag system possesses integrable dispersive deformations for special values of the parameters $\omega_k$, thus providing multi-component generalisations of examples from Theorem \ref{th2}. These deformations  are utilised in Section \ref{sec:vort} to construct dispersive regularisation  of a system governing  propagation of long nonlinear waves in a shear flow with piecewise linear velocity profile (for special values of vorticities $\omega_k$). The most general multi-component integrable system arising from the $2+1$ dimensional waterbag system (where we know the structure of dispersive deformations) corresponds to the case when $\omega_1=\dots =\omega_n=1$ and $\omega_{n+1}=\dots =\omega_{n+m}=-1$. 
Setting $u^{n+\alpha}=v^{\alpha}$ we obtain dispersionless equations
$$
u^i_t = u^iu^i_y + w_y, \quad v^{\alpha}_t = v^{\alpha}v^{\alpha}_y + w_y,\quad
w_x =\underset{k=1}{\overset{n}{\sum }}u_{y}^{k}-  \underset{\alpha =1}{\overset{m}{\sum }}v_{y}^{\alpha}.
$$
This system possesses integrable dispersive deformation
\begin{equation}\label{ILWn}
\begin{array}{l}
\vspace{2mm}
u^i_t = u^iu^i_y + w_y + \frac{\epsilon}{2}(v^1+\dots +v^m)_{yy}+\frac{\epsilon}{2}D^i_ju^j_{yy},\\ 
\vspace{2mm}
v^{\alpha}_t = v^{\alpha}v^{\alpha}_y + w_y + \frac{\epsilon}{2}(u^1+\dots +u^n)_{yy}+\frac{\epsilon}{2}D^{\alpha}_{\beta}v^{\beta}_{yy},\\
\triangle w = \frac{T + 1}{2}\bigg(\underset{k=1}{\overset{n}{\sum }}u_{y}^{k}-  \underset{\alpha =1}{\overset{m}{\sum }}v_{y}^{\alpha}\bigg),
\end{array}
\end{equation}
where $D^i_j$ and $D^{\alpha}_{\beta}$ are  the $n\times n$ and $m\times m$ skew-symmetric matrices with $1$'s and $-1$'s above/below the main diagonal. The corresponding Lax pair is
$$
\underset{\alpha =1}{\overset{m}\prod} (\epsilon\partial_y - v^{\alpha})T\psi = \underset{k =1}{\overset{n}\prod} (\epsilon\partial_y - u^k)\psi, \qquad
\epsilon\psi_t = \frac{\epsilon^2}{2}\psi_{yy} + \bigg(w + \frac{\epsilon}{2}\big(\underset{\alpha =1}{\overset{m}{\sum }}v_{y}^{\alpha}-\underset{k =1}{\overset{n}{\sum }}u_{y}^{k}\big)\bigg)\psi.
$$
We believe that  system (\ref{ILWn}) plays the role of a `master integrable system' in the classification of multi-component integrable equations with nonlocality of ILW type.

 Lax pairs for systems from Theorem \ref{th2} are  provided in the table below. They are obtained as quantisations of dispersionless Lax pairs \cite{Zakharov} (which are presented in the form that makes the quantisation procedure intuitively clear).

 \begin{center}
\begin{tabular}{| l | l | l |} \hline
 $Eqn$ & $Lax~pair$ &  $Dispersionless\ Lax\ pair$ \\ 
   & &    \\ \hline 
  &  &  \\
(\ref{ILW1}) & $2\epsilon T\psi_y = -2 \epsilon \psi_{y} +(v-u)T\psi + (u+v)\psi$ &  $e^{S_x}  = \frac{u+v-2S_{y}}{u-v+2S_{y}}$ \\
 & $\epsilon \psi_t= \epsilon^2 \psi_{yy} + (2w-\epsilon u_y) \psi$ &  $S_t=S_y^2+2w$ \\ 
   &  & \\ \hline
  &  & \\
(\ref{ILW2}) & $\epsilon T\psi_y = e^{-u}v\psi  - \epsilon e^{-u}\psi_y $ &  $e^{S_x}  = \frac{e^{-u}(v-S_y)}{S_y}$ \\
 & $ \psi_t= \frac{\epsilon}{2} \psi_{yy} +  (w - \frac{\epsilon}{2} u_y) \psi_y$ &  $S_t=\frac{1}{2}S_y^2+wS_y$ \\ 
  &  & \\ \hline
  &  & \\
(\ref{ILW3}) & $\epsilon T\psi_y = 2\epsilon e^{2u} \psi_y  - 2ve^{u}\sech u\ (T\psi+2\psi)$ &  $e^{S_x}  = 2e^{2u}\frac{S_y-2e^{-u}v\sech u}{S_y+2e^{u}v\sech u}$ \\
 & $ \psi_t= -\frac{\epsilon}{4} \psi_{yy} +  (w - \frac{\epsilon}{2} u_y) \psi_y$ &  $S_t  = -\frac{1}{4}S_y^2+wS_y$ \\ 
  &  & \\ \hline
  &  & \\
(\ref{ILW4}) & $T\psi = \frac{\epsilon^2}{4}\psi_{yy} - \epsilon u \psi_y + (u^2 - \frac{\epsilon}{2}u_y - \frac{\epsilon}{2}v_y - v^2)\psi $ &  $e^{S_x}  = \frac{1}{4}S_y^2-uS_y+u^2-v^2$ \\
& $\epsilon \psi_t= \frac{\epsilon^2}{4} \psi_{yy} + (2w - \epsilon u_y)\psi$  &  $S_t  = \frac{1}{4}S_y^2 + 2w$ \\ 
  &  & \\ \hline
    &  & \\
(\ref{ILW5}) & $T\psi = \frac{\epsilon^2}{4}\psi_{yy} - \epsilon u \psi_y + (u^2 - \frac{\epsilon}{2}(1 + i)u_y  - 2v)\psi$ &  $e^{S_x}  = \frac{1}{4}S_y^2-uS_y+u^2-2v$ \\
& $\epsilon \psi_t= \frac{\epsilon^2}{4} \psi_{yy} + (2w - \epsilon u_y)\psi$  &  $S_t= \frac{1}{4}S_y^2 + 2w $ \\ 
  &  & \\ \hline
  &  & \\
(\ref{ILW6}) & $T\psi = \frac{1}{2}\epsilon^2{e^{-2u}}\psi_{yy} - \frac{1}{2}\epsilon^2e^{-2u}(u_y + \frac{v_y}{\sqrt{v^2 + \alpha}})\psi_y - 2\epsilon ve^{-u} \psi_y - 2\alpha\psi $ &  $e^{S_x}  = \frac{1}{2}e^{-2u}S_y^2 -2ve^{-u}S_y-2\alpha$ \\
& $\psi_t= \frac{\epsilon}{4} \psi_{yy} + (w - \frac{\epsilon}{2} u_y)\psi_y$  &  $S_t  = \frac{1}{4}S_y^2 + wS_y$ \\ 
 &  &  \\ \hline
\end{tabular}
\end{center}

\section{Proof of Theorem \ref{th2}}
\label{sec:proof}

Our approach is based on a two-step procedure which can be summarised as follows:
\begin{itemize}
\item First, we consider dispersionless limit of system (\ref{ILW}) and require that the corresponding Haantjes tensor vanishes
\cite{FK}.  This gives  necessary conditions for integrability in the form of differential constraints for the coefficients $\alpha, \beta, \gamma, \phi, \psi, \eta$, leading to the two essentially different cases, see details below. 
\item Secondly, we require that all one-phase hydrodynamic reductions (simple waves) of the dispersionless limits obtained at the previous step  can be deformed  into reductions of the full dispersive system (\ref{ILW}). This approach, which can be seen as a $2+1$ dimensional extension of the deformation procedure developed by the Dubrovin school  in $1+1$ dimensions \cite{Dub1, Dub2},  has been proposed in \cite{FerM, FMN} and successfully adapted to the classification of integrable systems in $2+1$ dimensions, see \cite{NF, HN, FNR, FNR1}. Applied to system (\ref{ILW}), it results in further constraints for the dispersionless limit and $\epsilon$-corrections, eventually leading to the six cases of Theorem \ref{th2}. 
\end{itemize}

Let us go through the above scheme in some more detail. The dispersionless limit of system (\ref{ILW}) is
\begin{equation}\label{ILWdis}
\begin{array}{l}
\vspace{2mm}
u_t=\alpha(u,v,w)u_y+\beta(u,v,w)v_y+\gamma(u,v,w)w_y,\\ 
\vspace{2mm}
v_t=\phi(u,v,w)u_y+\psi(u,v,w)v_y+\eta(u,v,w)w_y,\\
w_x=u_y.
\end{array}
\end{equation}
Introducing the column vector $U=(u^1, u^2, u^3)^T=(u, v, w)^T$, we can represent system  (\ref{ILWdis}) in matrix form,
\begin{equation}\label{mat}
AU_t+BU_x+CU_y=0,
\end{equation}
where the $3\times 3$ matrices $A, B$ and $C$ are as follows:
$$
A=\left(
\begin{array}{ccc}
-1&0&0\\
0&-1&0\\
0&0&0\end{array}\right), \qquad B=\left(
\begin{array}{ccc}
0&0&0\\
0&0&0\\
0&0&-1\end{array}\right), \qquad C=\left(
\begin{array}{ccc}
\alpha&\beta&\gamma\\
\phi&\psi&\eta\\
1&0&0\end{array}\right).          
$$
The dispersion relation of system (\ref{mat}) is defined by the formula $\det (\lambda A+\mu B+C)=0$, which gives a rational cubic curve:
\begin{equation}\label{disr}
\mu((\alpha-\lambda)(\psi-\lambda)-\beta \phi)+\gamma(\phi-\lambda)-\beta\eta=0.
\end{equation}
In what follows we will assume that the dispersion relation defines an irreducible cubic.

Let us introduce the matrix $V=(B+pA)^{-1}(C+qA)$ where $p$ and $q$ are arbitrary parameters. It was demonstrated in \cite{FK} that the necessary condition for integrability of the dispersionless limit (\ref{ILWdis}) is the vanishing of the Haantjes tensor \cite{Haantjes} of the matrix $V$, identically in the parameters $p, q$. Recall that, given a matrix $V$, its   Haantjes tensor is defined by the formula
$$
H^i_{jk}=N^i_{pr}V^p_jV^r_k-N^p_{jr}V^i_pV^r_k-N^p_{rk}V^i_pV^r_j+N^p_{jk}V^i_rV^r_p,
$$ 
where
$$
N^i_{jk}=V^p_j\partial_{u^p}V^i_k-V^p_k\partial_{u^p}V^i_j-V^i_p(\partial_{u^j}V^p_k-\partial_{u^k}V^p_j)
$$
is the Nijenhuis tensor. The requirement of vanishing of the Haantjes tensor imposes strong constraints on the coefficients of system (\ref{ILWdis}). One of them is the relation
$$
\gamma \eta_w=\eta \gamma_w,
$$
which splits the further analysis into  two cases:

{\bf Case 1:}  $\gamma=0$.

{\bf Case 2:}  $\gamma \ne0$. In this case we have $\eta =s(u, v)\gamma$. Utilising the transformation freedom $v\to f(u, v)$, we can set  $\eta=0$.

Let us discuss Case 1 in some more detail (Case 2 can be treated analogously). If $\gamma=0$, the vanishing of the Haantjes tensor implies further constraints such as
$$
\beta_w=\phi_w=\eta_w=(\alpha-\psi)_w=0, 
$$
etc. In particular, $\beta=b(u, v)$, and utilising the transformation freedom $v\to f(u, v)$, we can set  $\beta=1$;
note that $\beta$ cannot vanish, otherwise,  the dispersion relation (\ref{disr}) would  become reducible.  
Further analysis of the Haantjes tensor leads to the two subcases.

Subcase 1.1:
\begin{equation*}
\begin{array}{l}
\vspace{2mm}
u_t=-\frac{1}{2}(\alpha+u\alpha')u_y+v_y,\\ 
\vspace{2mm}
v_t=(-\frac{v^2}{u^2}+au^2-v \alpha'-\frac{1}{4}u^2{\alpha'}^2)u_y+(2\frac{v}{u}-\frac{1}{2}\alpha+\frac{1}{2}u\alpha')v_y+bu w_y,\\
w_x=u_y,
\end{array}
\end{equation*}
where $\alpha (u)$ is an arbitrary function  and $a, b$ are arbitrary constants. The corresponding Haantjes tensor vanishes identically.
Note that the arbitrary function $\alpha(u)$ can be eliminated by the substitution $v\to v+\frac{1}{2}u\alpha(u)$, leading to the simplified form
\begin{equation}\label{IL3}
\begin{array}{l}
\vspace{2mm}
u_t=v_y,\\ 
\vspace{2mm}
v_t=(-\frac{v^2}{u^2}+au^2)u_y+\frac{v}{u}v_y+bu w_y,\\
w_x=u_y,
\end{array}
\end{equation}
(particular values of constants $a, b$ are not important, we will normalise them at a later stage).

Subcase 1.2:
\begin{equation*}
\begin{array}{l}
\vspace{2mm}
u_t=(cw+\alpha')u_y+v_y,\\ 
\vspace{2mm}
v_t=(-\frac{1}{2}cv^2\eta'-cv\alpha \eta'-\frac{1}{2}c\alpha^2 \eta'
-cv\eta \alpha'-c\alpha \eta \alpha'-{\alpha'}^2)u_y
+(cw-cv\eta-c\alpha \eta-\alpha')v_y+c(v+\alpha) w_y,\\
w_x=u_y,
\end{array}
\end{equation*}
where $\alpha (u)$ is an arbitrary function, $c$ is an arbitrary constant and the function $\eta(u)$ satisfies the second-order ODE, $\eta''=c\eta \eta'$. The corresponding Haantjes tensor  vanishes identically. Again, the arbitrary function $\alpha(u)$ can be eliminated by the substitution $v\to v-\alpha(u)$, leading to the simplified form
\begin{equation}\label{IL4}
\begin{array}{l}
\vspace{2mm}
u_t=cwu_y+v_y,\\ 
\vspace{2mm}
v_t=-\frac{1}{2}cv^2\eta'u_y
+c(w-v\eta)v_y+cv w_y,\\
w_x=u_y.
\end{array}
\end{equation}
Note that one can scale the constant to $c=1$, then ODE $\eta''=\eta \eta'$ leads to the two further  subcases: $\eta=1$ and $\eta =-2\tanh u$.

To summarise, based on the requirement of vanishing of the Haantjes tensor only, in Case 1 we have found three candidates for dispersionless limits of integrable systems (\ref{ILW}), namely, systems (\ref{IL3}) and (\ref{IL4}) (for two different forms of $\eta$). The next step of the classification  is to consider one-phase reductions of these dispersionless systems and require that they can be deformed into reductions of the corresponding dispersive systems  (\ref{ILW}).

This step is computationally intense, the details are as follows. Recall that one-phase reductions (simple waves) of system (\ref{mat}) are defined by the formula $U=U(R)$ where the `phase' $R$ satisfies a pair of compatible equations 
\begin{equation}\label{R}
R_t=\lambda(R) R_y, \quad R_x=\mu(R) R_y.
\end{equation} 
Substituting this ansatz into (\ref{mat}) we obtain the relation
$$
(\lambda A+\mu B+C)U'=0
$$
where $U'=dU/dR$. Thus, $U'$ is the eigenvector of  the matrix $\lambda A+\mu B+C$, while the characteristic speeds $\lambda, \mu$ satisfy the dispersion relation,  $\det(\lambda A+\mu B+C)=0$. Let us look for  solutions of  system (\ref{ILW}) in the form
\begin{equation}\label{U}
U=U(R)+\epsilon(\dots)+\epsilon^2(\dots)+\dots,
\end{equation}
where dots at $\epsilon^n$ denote differential polynomials in $y$-derivatives of $R$ of the total degree $n$, whose coefficients are some functions of $R$ (expansions of this kind do not terminate in general).  
Similarly, we assume that  equations (\ref{R}) are also deformed,
\begin{equation}\label{Rdef}
R_t=\lambda(R) R_y+\epsilon(\dots)+\epsilon^2(\dots)+\dots, \quad R_x=\mu(R) R_y+\epsilon(\dots)+\epsilon^2(\dots)+\dots;
\end{equation}
here dots at $\epsilon^n$ denote differential polynomials in $y$-derivatives of $R$ of the total degree $n+1$.  Let us substitute (\ref{U}) into (\ref{ILW}). Using (\ref{Rdef}) and requiring that  the terms at the same powers of $\epsilon$ vanish identically, we can explicitly reconstruct the  expansions (\ref{U}) and (\ref{Rdef}), where all terms denoted by dots can be uniquely expressed  in terms of $U(R), \lambda(R), \mu(R)$, and the  $\epsilon$-terms in equation (\ref{ILW}). Furthermore, by looking at terms of the order not higher than $\epsilon^2$, we get strong necessary conditions for the $\epsilon$-terms in (\ref{ILW}) which, in this classification problem, turn out to be also sufficient (this can be demonstrated by directly constructing the Lax pairs).  

Applied to dispersionless limit (\ref{IL3}), the above procedure  leads to case (\ref{ILW1}) of Theorem \ref{th2} (up to the change of variables $v\to uv$ and suitable rescalings).

Similarly, applied to dispersionless limit (\ref{IL4}), the above procedure  leads to cases (\ref{ILW2}) and (\ref{ILW3})  of Theorem \ref{th2} (which correspond to $\eta=1$ and $\eta =-2\tanh u$, respectively). 

Similarly, Case 2 leads to systems (\ref{ILW4}), (\ref{ILW5}) and (\ref{ILW6}).
\qed

\section{Waterbag system in 2+1 dimensions}
\label{sec:wb}

Several equations  appearing in Theorems \ref{th1}-\ref{th2} can be obtained, as special cases, from the $2+1$ dimensional dispersionless integrable system
\begin{equation}
u^{i}_{t}=u^iu^{i}_{y}+w_y,\qquad
w_{x}=\underset{k=1}{\overset{N}{\sum}}\frac{1}{\omega_{k}}u^{k}_{y}, \label{wb}
\end{equation}
$i=1, \dots, N$, which possesses the Lax pair
$$
e^{S_{x}}=\underset{k=1}{\overset{N}{\prod}}\ (S_{y}-u^k)^{\frac{1}{\omega_k}},\text{
\ \ }S_{t}=\frac{1}{2}S_{y}^{2}+w;
$$
here $\omega_{k}$ are arbitrary constants. In the $1+1$ dimensional limit,  $y=-x$,  system (\ref{wb}) reduces to 
\begin{equation}
u^{i}_{t}+u^iu^i_{x}-\underset{k=1}{\overset{N}{\sum}}\frac{1}{\omega_{k}}u^k_x=0,
\label{CEP}
\end{equation}
 the so-called `waterbag' reduction of the  Benney chain \cite{GibTsa96}. It was shown in  \cite{CEGP} that a special case of system (\ref{CEP}) describes propagation of long nonlinear waves in a  shear flow with piecewise linear velocity profile, see Section \ref{sec:vort}.
 
Dispersive integrable extension of system (\ref{wb}) is not known (for generic  $\omega_k$).  However, for special values $\omega_{k}=\pm1$, the corresponding dispersive  extensions are associated with Lax pairs of the form
$$
P(\epsilon \partial_y) T\psi=Q(\epsilon \partial_y)\psi, 
\qquad \epsilon \psi_{t}=\frac{\epsilon^2}{2}\psi_{yy}+w\psi,
$$
where $P$ and $Q$ are differential operators whose coefficients depend on $u$'s. Below we discuss some special cases of this construction.

\subsection{Case $N=1, \ \omega_1=1$}

The corresponding system (\ref{wb}) takes the form
$$
u_{t}=uu_{y}+w_y, \quad w_x=u_y,
$$
with the  dispersionless Lax pair 
$$
e^{S_{x}}=S_{y}-u,\text{ \ \ }S_{t}=\frac{1}{2}S_{y}^{2}+w.
$$
This system possesses dispersive deformation (the first case of Theorem 1):
$$
u_{t}=uu_{y}+w_y, \quad \triangle w=\frac{T+1}{2}u_{y},
$$
with the Lax pair
$$
T\psi=\epsilon \psi_y-u\psi, \qquad
\epsilon \psi_t=\frac{\epsilon^2}{2}\psi_{yy}+(w-\frac{\epsilon}{2}u_y)\psi.
$$

\subsection{Case $N=2, \ \omega_1=1, \  \omega_2=-1$}

The corresponding system (\ref{wb}) takes the form 
\begin{equation}\label{N=2}
u_{t}=uu_y+w_y,\text{ \ }v_{t}=vv_y+w_y,\text{ \ }w_{x}=(u-v)_{y},
\end{equation}
with the dispersionless  Lax pair 
$$
e^{S_{x}}=\frac{S_{y}-u}{S_{y}-v},\text{ \ \ }S_{t}=\frac{1}{2}S_{y}^{2}+w.
$$
System (\ref{N=2})  possesses the following dispersive deformation:
\begin{equation}\label{disN=2}
u_{t}=uu_{y}+w_{y}+\frac{\epsilon}{2}v_{yy},\text{ \ \ }v_{t}=vv_{y}
+w_{y}+\frac{\epsilon}{2}u_{yy},
\quad
\triangle w=\frac{T+1}{2}(u-v)_{y},
\end{equation}
with the Lax pair
\begin{equation}\label{LaxdisN=2}
(\epsilon \partial_y-v)T\psi=(\epsilon \partial_y-u)\psi,  \qquad
\epsilon \psi_t=\frac{\epsilon^2}{2}\psi_{yy}+(w+\frac{\epsilon}{2}(v-u)_y)\psi.
\end{equation}
Note that system (\ref{disN=2}) coincides with (\ref{ILW1}) in the variables  $\hat{u}=u-v$, $\hat{v}=u+v$ (up to unessential scaling factors). System (\ref{disN=2}) is equivalent to the differential-difference Davey-Stewartson (DS)  equation. To see this, we rewrite Lax pair  (\ref{LaxdisN=2}) in equivalent form,
\begin{equation}\label{NLS1}
T\psi=(1+\frac{1}{\epsilon}p\partial_y^{-1}q)\psi,  \qquad
\epsilon \psi_t=\frac{\epsilon^2}{2}\psi_{yy}+\tilde w \psi,
\end{equation}
where the new variables $p, q, \tilde w$ are connected to $u, v, w$ by the formulae
$$
u=\epsilon \frac{p_y}{p}-pq,   \quad v=\epsilon \frac{p_y}{p}, \quad w=\tilde w-\frac{\epsilon}{2}(v-u)_y.
$$
The compatibility conditions of Lax pair (\ref{NLS1}) result in the differential-difference DS equation,
\begin{equation}\label{NLS2}
\epsilon p_t=\frac{\epsilon^2}{2}p_{yy}+pT\tilde w, \quad   -\epsilon q_t=\frac{\epsilon^2}{2}q_{yy}+q\tilde w, \quad \triangle \tilde w=-(pq)_y.
\end{equation}

\noindent {\bf Remark.} Lax pair (\ref{NLS1}) and  system (\ref{NLS2}) possess  natural multi-component extensions:
\begin{equation}\label{NLS3}
T\psi=(1+\frac{1}{\epsilon}\sum_{k}p^k\partial_y^{-1}q^k)\psi,  \qquad
\epsilon \psi_t=\frac{\epsilon^2}{2}\psi_{yy}+\tilde w \psi,
\end{equation}
and
\begin{equation}\label{NLS4}
\epsilon p^i_t=\frac{\epsilon^2}{2}p^i_{yy}+p^iT\tilde w, \quad   -\epsilon q^i_t=\frac{\epsilon^2}{2}q^i_{yy}+q^i\tilde w, \quad \triangle \tilde w=-(\sum_{k} p^kq^k)_y,
\end{equation}
respectively. This multi-component differential-difference DS system is apparently new. Applying the Madelung transformation,
$$
p^i=\sqrt {\eta^i}\ e^{\frac{1}{\epsilon} \int u^i dy},\quad q^i=\sqrt {\eta^i}\ e^{-\frac{1}{\epsilon} \int u^i dy},
$$
one can rewrite system (\ref{NLS4}) in the Hasimoto form,
$$
\eta^i_t=(u^i\eta^i)_y+\eta^i\triangle \tilde w, \quad u^i_t=u^iu^i_y+\frac{T+1}{2}\tilde w_y+\frac{\epsilon^2}{4}\left(\frac{\eta^i_{yy}}{\eta^i}-\frac{1}{2}\frac{(\eta_y^i)^2}{(\eta^i)^2} \right)_y, \quad \triangle \tilde w=-(\sum_{k} \eta^k)_y.
$$
Introducing the new variables $v^i$ by the formula $v^i=u^{i}-\frac{\epsilon}{2}(\ln \eta
^{i})_{y}$ we obtain  the equivalent Kaup-Broer form:
$$
\eta^i_t=(v^i\eta^i)_y+\eta^i\triangle \tilde w+\frac{\epsilon}{2}\eta^i_{yy}, \quad v^i_t=v^iv^i_y+\frac{T+1}{2}\tilde w_y-\frac{\epsilon}{2}\left(v^i_y+\triangle \tilde w \right)_y, \quad \triangle \tilde w=-(\sum_{k} \eta^k)_y.
$$

\subsection{Case $N=2, \ \omega_1=-\omega_2$}

Setting $\omega_1=-\omega_2=\omega$, one can rescale this case back to the previous one by setting $x=\omega \tilde x$.  However, our goal here is to investigate the limit $\omega \to 0$. Thus, system (\ref{disN=2}) and its Lax pair (\ref{LaxdisN=2}) assume the form
\begin{equation}\label{disN=21}
u_{t}=uu_{y}+w_{y}+\frac{\epsilon}{2}v_{yy},\text{ \ \ }v_{t}=vv_{y}
+w_{y}+\frac{\epsilon}{2}u_{yy},
\quad
\frac{\tilde T-1}{\epsilon} w=\frac{\tilde T+1}{2}(u-v)_{y},
\end{equation}
and
\begin{equation}\label{LaxdisN=21}
(\epsilon \partial_y-v)\tilde T\psi=(\epsilon \partial_y-u)\psi,  \qquad
\epsilon \psi_t=\frac{\epsilon^2}{2}\psi_{yy}+(w+\frac{\epsilon}{2}(v-u)_y)\psi,
\end{equation}
respectively; here $\tilde T=e^{\epsilon \partial_{\tilde x}}=e^{\epsilon \omega \partial_x}$.
Setting $u=v-\omega \eta $, we can rewrite (\ref{disN=21}) and (\ref{LaxdisN=21}) in the form
$$
\eta_t=(v\eta)_y-\omega \eta \eta_y-\frac{\epsilon}{2}\eta_{yy}, \qquad v_t=vv_y+w_y+\frac{\epsilon}{2}v_{yy}-\frac{\epsilon \omega}{2}\eta_{yy}, \qquad \frac{\tilde T-1}{\epsilon \omega} w=-\frac{\tilde T+1}{2}\eta_{y},
$$
and
$$
\epsilon (\epsilon \partial_y-v)\frac{\tilde T-1}{\epsilon \omega}\psi=\eta \psi,  \qquad
\epsilon \psi_t=\frac{\epsilon^2}{2}\psi_{yy}+(w+\frac{\epsilon \omega}{2}\eta_y)\psi,
$$
respectively. In the limit $\omega \to 0$ we have $\frac{\tilde T-1}{\epsilon \omega}\to \partial_x$, which results in the system
$$
\eta_t=(v\eta)_y-\frac{\epsilon}{2}\eta_{yy}, \qquad v_t=vv_y+w_y+\frac{\epsilon}{2}v_{yy}, \qquad  w_x=-\eta_{y},
$$
and its Lax pair, 
$$
\epsilon (\epsilon \partial_y-v)\psi_x=\eta \psi,  \qquad
\epsilon \psi_t=\frac{\epsilon^2}{2}\psi_{yy}+w\psi,
$$
see \cite{Zakharov}, Section 3.

\subsection{Case $N=2, \ \omega_1= \omega_2=1$}

The corresponding system (\ref{wb}) takes the form 
\begin{equation}\label{11}
u_{t}=uu_y+w_y,\text{ \ }v_{t}=vv_y+w_y,\text{ \ }w_{x}=(u+v)_{y},
\end{equation}
with the dispersionless  Lax pair 
$$
e^{S_{x}}=(S_{y}-u)(S_{y}-v),\text{ \ \ }S_{t}=\frac{1}{2}S_{y}^{2}+w.
$$
System (\ref{11}) possesses dispersive deformation
$$
u_{t}=uu_{y}+w_{y}+\frac{\epsilon}{2}v_{yy},\text{ \ \ }v_{t}=vv_{y}
+w_{y}-\frac{\epsilon}{2}u_{yy},
\quad
\triangle w=\frac{T+1}{2}(u+v)_{y},
$$
with the Lax pair
$$
T\psi=(\epsilon \partial_y-u)(\epsilon \partial_y-v)\psi,  \qquad
\epsilon \psi_t=\frac{\epsilon^2}{2}\psi_{yy}+(w-\frac{\epsilon}{2}(u+v)_y)\psi.
$$
Up to elementary changes of variables, this case is equivalent to systems (\ref{ILW4}), (\ref{ILW5}):  this can be seen from the structure of the corresponding Lax pairs.  
Interestingly, system (\ref{11}) possesses yet another (third-order) dispersive deformation, 
$$
u_{t}=uu_{y}+w_{y}-\frac{\epsilon^2}{4}\frac{u_{yyy}+v_{yyy}}{u-v},\qquad v_{t}=vv_{y}
+w_{y}-\frac{\epsilon^2}{4}\frac{u_{yyy}+v_{yyy}}{v-u},
\qquad
\triangle w=\frac{T+1}{2}(u+v)_{y},
$$
associated with the `symmetrised' Lax pair,
$$
T\psi=(\epsilon^2 \partial^2_y-\epsilon (u+v) \partial_y+uv-\frac{\epsilon}{2}(u+v)_y)\psi,  \qquad
\epsilon \psi_t=\frac{\epsilon^2}{2}\psi_{yy}+(w-\frac{\epsilon}{2}(u+v)_y)\psi.
$$
Both dispersive deformations are clearly Miura-equivalent.


\subsection{Case $N=n, \ \omega_1= \dots=\omega_n=1$}
\label{sec:cont}

Let us begin with the $3$-component case,
\begin{equation}\label{11}
u^i_{t}=u^iu^i_y+w_y, \qquad w_{x}=(u^1+u^2+u^3)_{y},
\end{equation}
$i=1, 2, 3$, with the dispersionless  Lax pair 
$$
e^{S_{x}}=(S_{y}-u^1)(S_{y}-u^2)(S_y-u^3),\text{ \ \ }S_{t}=\frac{1}{2}S_{y}^{2}+w.
$$
 System (\ref{11}) possesses dispersive deformation
\begin{align*}
u^1_t &= u^1u^1_y + w_y + \frac{\epsilon}{2}(u^2 + u^3)_{yy},\\
u^2_t &= u^2u^2_y + w_y + \frac{\epsilon}{2}(-u^1+u^3)_{yy},\\
u^3_t &= u^3u^3_y + w_y + \frac{\epsilon}{2}(-u^1 - u^2)_{yy},
\end{align*}
$$
\triangle w = \frac{T + 1}{2}(u^1 + u^2 + u^3)_y,
$$
with the Lax pair
$$
T\psi = (\epsilon\partial_y - u^1)(\epsilon\partial_y - u^2)(\epsilon\partial_y - u^3)\psi, \qquad
\epsilon\psi_t = \frac{\epsilon^2}{2}\psi_{yy} + \big(w - \frac{\epsilon}{2}(u^1 + u^2 + u^3)_y\big)\psi.
$$
The general $n$-component version is as follows:
$$
u^i_t = u^iu^i_y + w_y + \frac{\epsilon}{2}D^i_ju^j_{yy},
$$
$$
\triangle w = \frac{T + 1}{2}\sum u^k_y,
$$
where $D^i_j$ is the $n\times n$ skew-symmetric matrix with $1$'s and $-1$'s above/below the main diagonal.
The corresponding Lax pair is
$$
T\psi = \prod (\epsilon\partial_y - u^k)\psi, \qquad
\epsilon\psi_t = \frac{\epsilon^2}{2}\psi_{yy} + \big(w - \frac{\epsilon}{2}\sum u^k_y\big)\psi.
$$

\subsection{Case $N=2n, \ \omega_1=\dots =\omega_n=1, \ \   \omega_{n+1}=\dots=\omega_{2n}=-1$} 
\label{sec:2n}

The corresponding system (\ref{wb}) takes the form (we set $v^i=u^{n+i}$):
$$
u^i_{t}=u^iu^i_y+w_{y},\quad v^i_{t}=v^iv^i_y+w_{y},\quad { \ }w_{x}
=\underset{k=1}{\overset{n}{\sum}}(u^k-v^k)_{y}.
$$
It possesses  dispersionless  Lax pair
$$
e^{S_{x}}=\underset{k=1}{\overset{n}{
{\displaystyle\prod}
}}  \frac{S_{y}-u^{k}}{S_{y}-v^{k}}, \quad S_{t}=\frac{1}{2}S_{y}^{2}+w,
$$
which has dispersive extension of the form
$$
 P(\epsilon \partial_y) T\psi=Q(\epsilon \partial_y) \psi, \quad \epsilon \psi_t=\frac{\epsilon^2}{2}\psi_{yy}+ w \psi;
 $$
here $P$ and $Q$ are  differential operators of degree $n$ whose coefficients depend on $u$'s and $v$'s. This Lax pair and the associated dispersive system are Miura-equivalent to 
(\ref{NLS3}), (\ref{NLS4}). As an example let us consider the case $N=4$. 
The Lax pair 
$$
(\epsilon\partial_y - v^1)(\epsilon\partial_y - v^2) T\psi =(\epsilon\partial_y - u^1)(\epsilon\partial_y - u^2)\psi,
$$
$$
\epsilon\psi_t = \frac{\epsilon^2}{2}\psi_{yy} + \big(w + \frac{\epsilon}{2}(v^1 + v^2 - u^1 - u^2)_y\big)\psi,
$$
results in the second-order  deformation,
\begin{align*}
u^1_t &= u^1 u^1_y + w_y + \frac{\epsilon}{2}(v^1 + v^2+ u^2)_{yy},\\
u^2_t &= u^2 u^2_y + w_y + \frac{\epsilon}{2}(v^1 + v^2 - u^1)_{yy},\\ 
v^1_t &= v^1 v^1_y + w_y + \frac{\epsilon}{2}(u^1 + u^2 + v^2)_{yy},\\
v^2_t &= v^2 v^2_y + w_y + \frac{\epsilon}{2}(u^1 + u^2 - v^1)_{yy} ,
\end{align*}
$$
\triangle w = \frac{T + 1}{2}(u^1 + u^2 - v^1 - v^2)_y.
$$
The `symmetrised' Lax pair,
$$
\big((\epsilon\partial_y - v^1)(\epsilon\partial_y - v^2) + (\epsilon\partial_y - v^2)(\epsilon\partial_y - v^1)\big)T\psi = \big((\epsilon\partial_y - u^1)(\epsilon\partial_y - u^2) + (\epsilon\partial_y - u^2)(\epsilon\partial_y - u^1)\big)\psi,
$$
$$
\epsilon\psi_t = \frac{\epsilon^2}{2}\psi_{yy} + \big(w + \frac{\epsilon}{2}(v^1 + v^2 - u^1 - u^2)_y\big)\psi,
$$
results in the third-order deformation,
\begin{align*}
u^1_t &= u^1 u^1_y + w_y + \frac{\epsilon}{2}(v^1_{yy} + v^2_{yy}) - \frac{\epsilon^2}{4}\frac{u^1_{yyy} + u^2_{yyy}}{u^1 - u^2},\\
u^2_t &= u^2 u^2_y + w_y + \frac{\epsilon}{2}(v^1_{yy} + v^2_{yy}) - \frac{\epsilon^2}{4}\frac{u^1_{yyy} + u^2_{yyy}}{u^2 - u^1},\\
v^1_t &= v^1 v^1_y + w_y + \frac{\epsilon}{2}(u^1_{yy} + u^2_{yy}) - \frac{\epsilon^2}{4}\frac{v^1_{yyy} + v^2_{yyy}}{v^1 - v^2},\\
v^2_t &= v^2 v^2_y + w_y + \frac{\epsilon}{2}(u^1_{yy} + u^2_{yy}) - \frac{\epsilon^2}{4}\frac{v^1_{yyy} + v^2_{yyy}}{v^2 - v^1},
\end{align*}
$$
\triangle w = \frac{T + 1}{2}(u^1 + u^2 - v^1 - v^2)_y.
$$
Both second-order and third-order deformations are Miura-equivalent. 

The general $2n$-component version is as follows:
\begin{align*}
u^i_t &= u^iu^i_y + w_y + \frac{\epsilon}{2}(v^1+\dots +v^n)_{yy}+\frac{\epsilon}{2}D^i_ju^j_{yy},\\
v^i_t &= v^iv^i_y + w_y + \frac{\epsilon}{2}(u^1+\dots +u^n)_{yy}+\frac{\epsilon}{2}D^i_jv^j_{yy}
\end{align*}
$$
\triangle w = \frac{T + 1}{2}\sum (u^k-v^k)_y,
$$
where $D^i_j$ is the $n\times n$ skew-symmetric matrix with $1$'s and $-1$'s above/below the main diagonal.
The corresponding Lax pair is
$$
\prod (\epsilon\partial_y - v^k)T\psi = \prod (\epsilon\partial_y - u^k)\psi, \qquad
\epsilon\psi_t = \frac{\epsilon^2}{2}\psi_{yy} + \big(w + \frac{\epsilon}{2}\sum (v^k-u^k)_y\big)\psi.
$$

\subsection{Case $N=2n, \ \omega_{n+i}=-\omega_i$}

Setting $u^{n+i}=v^i$ one can write the corresponding equations in the form
\begin{equation}\label{shear}
u_{t}^{i}=u^{i}u_{y}^{i}+w_{y},\qquad v_{t}^{i}=v^{i}v_{y}^{i}+w_{y}, \qquad w_{x}=\underset{k=1}{\overset{n}{\sum 
}}\frac{u_{y}^{k}-v^k_y}{\omega_{k}}.
\end{equation}
This system possesses  dispersionless Lax pair
\begin{equation*}
e^{S_{x}}=\underset{k=1}{\overset{n}{\prod }}\ \left(\frac{S_{y}-u^{k}}{S_y-v^k}\right)^{\frac{1}{\omega_k}
},\quad S_{t}=\frac{1}{2}S_{y}^{2}+w, 
\end{equation*}
equivalently,
\begin{equation*}
S_{x}=\underset{k=1}{\overset{n}{\sum }}\frac{1}{\omega _{k}}\ln 
\frac{S_{y}-u^{k}}{S_{y}-v^{k}}, \qquad  S_{t}=\frac{1}{2}S_{y}^{2}+w.
\end{equation*}
System (\ref{shear}) can be viewed as a $2+1$ dimensional generalisation of the $1+1$ dimensional system governing shear flow in $n$-layered fluid, see \cite{CEGP} and Section \ref{sec:vort} of this paper. 
Setting  $u^{k}=v^{k}-\omega
_{k}\eta ^{k}$, we can rewrite (\ref{shear}) in the form
\begin{equation}\label{shear1}
v_{t}^{i}=v^{i}v_{y}^{i}+w_{y}, \qquad  \eta^i_t=(v^i\eta^i)_y-\omega_i\eta^i\eta^i_y,     
\qquad w_{x}+\underset{k=1}{\overset{n}{\sum}} \eta^k_y=0.
\end{equation}
Taking the limit  $\omega_k\to 0$ we obtain
\begin{equation*}
S_{x}=\underset{k=1}{\overset{n}{\sum }}\frac{1}{\omega _{k}}\ln \frac{
S_{y}-v^{k}+\omega _{k}\eta ^{k}}{S_{y}-v^{k}}=\underset{k=1}{\overset{n}{
\sum }}\frac{1}{\omega _{k}}\ln \left( 1+\omega _{k}\frac{\eta ^{k}}{
S_{y}-v^{k}}\right) \approx \underset{k=1}{\overset{n}{\sum }}\frac{\eta ^{k}
}{S_{y}-v^{k}},
\end{equation*}
which gives the Lax pair
\begin{equation*}
S_{x}=\underset{k=1}{\overset{n}{\sum }}\frac{\eta ^{k}}{S_{y}-v^{k}},\qquad S_{t}=\frac{1}{2}S_{y}^{2}+w.
\end{equation*}
This Lax pair governs the $2+1$ dimensional  generalisation of $n$-layer Benney system discussed  in \cite{Zakharov}, 
\begin{equation*}
v_{t}^{i}=v^{i}v_{y}^{i}+w_{y}, \qquad  \eta^i_t=(v^i\eta^i)_y,     
\qquad w_{x}+\underset{k=1}{\overset{n}{\sum}} \eta^k_y=0,
\end{equation*}
which is the limit of (\ref{shear1}) as $\omega_k\to 0$.

\subsection{Case $N=n+m, \ \omega_{1}=\dots =\omega_n=1, \ \omega_{n+1}=\dots =\omega_{n+m}=-1$} 

This is the most general case where we know the structure of dispersive deformations.
Setting $u^{n+\alpha}=v^{\alpha}$ one can write the corresponding equations in the form
\begin{equation}\label{shear}
u_{t}^{i}=u^{i}u_{y}^{i}+w_{y},\qquad v_{t}^{\alpha}=v^{\alpha}v_{y}^{\alpha}+w_{y}, \qquad w_{x}=\underset{k=1}{\overset{n}{\sum }}u_{y}^{k}-  \underset{\alpha =1}{\overset{m}{\sum }}v_{y}^{\alpha}.
\end{equation}
($n$ variable $u^i$, $m$ variables $v^{\alpha}$). This system possesses integrable dispersive deformation
\begin{align*}
u^i_t &= u^iu^i_y + w_y + \frac{\epsilon}{2}(v^1+\dots +v^m)_{yy}+\frac{\epsilon}{2}D^i_ju^j_{yy},\\
v^{\alpha}_t &= v^{\alpha}v^{\alpha}_y + w_y + \frac{\epsilon}{2}(u^1+\dots +u^n)_{yy}+\frac{\epsilon}{2}D^{\alpha}_{\beta}v^{\beta}_{yy},
\end{align*}
$$
\triangle w = \frac{T + 1}{2}\bigg(\underset{k=1}{\overset{n}{\sum }}u_{y}^{k}-  \underset{\alpha =1}{\overset{m}{\sum }}v_{y}^{\alpha}\bigg),
$$
where $D^i_j$ and $D^{\alpha}_{\beta}$ are  $n\times n$ and $m\times m$ skew-symmetric matrices as in Section \ref{sec:2n}. The corresponding Lax pair is
$$
\underset{\alpha =1}{\overset{m}\prod} (\epsilon\partial_y - v^{\alpha})T\psi = \underset{k =1}{\overset{n}\prod} (\epsilon\partial_y - u^k)\psi, \qquad
\epsilon\psi_t = \frac{\epsilon^2}{2}\psi_{yy} + \bigg(w + \frac{\epsilon}{2}\big(\underset{\alpha =1}{\overset{m}{\sum }}v_{y}^{\alpha}-\underset{k =1}{\overset{n}{\sum }}u_{y}^{k}\big)\bigg)\psi.
$$

\section{Waterbag system and a  shear flow with piecewise linear velocity profile}
\label{sec:vort}

In this section we discuss a quasilinear  system governing  multilayer shear flow with  piecewise linear  velocity profile and constant vorticity within each layer \cite{CEGP}:
\begin{equation}\label{shear2}
u_{t}^{i}+u^{i}u_{x}^{i}+gh_x=0,\qquad v_{t}^{i}+v^{i}v_{x}^{i}+gh_x=0, \qquad h=\underset{k=1}{\overset{n}{\sum 
}}\frac{v^{k}-u^k}{\omega_{k}},
\end{equation}
where $v^i$ and $u^i$ are the velocities at the upper/lower boundary of the $i$-th layer and $\omega_i$ is the constant vorticity. It is known  that system (\ref{shear2}) is strictly hyperbolic, Hamiltonian and possesses Riemann invariants \cite{CEGP}. System (\ref{shear2})  can be viewed as an extension of Zakharov's multilayer reduction \cite{Zakharov2} of the Benney system.
 It has also appeared in the context of symmetry constraints of the dKP hierarchy \cite{BK}. We do not impose the continuity condition $v^i=u^{i+1}$, generally allowing jumps in horisontal velocities between layers. Such flows could model oceanic and industrial multilayer flows with strongly sheared currents.

One can see that system (\ref{shear2}) can be obtained from the $2+1$ dimensional waterbag system (\ref{shear}) by dimensional reduction $y=-x$. The results of Section \ref{sec:wb} allow one to construct dispersive regularisations for some special cases of system (\ref{shear2}). 
As the simplest example, let us consider the one-layer case, which is already nontrivial. Setting $u^1=u,\ v^1=v$ and normalising $g=\omega_1=1$ we obtain the system
\begin{equation}\label{1l}
u_{t}+uu_{x}+h_x=0,\qquad v_{t}+vv_{x}+h_x=0, \qquad h=v-u.
\end{equation}
System (\ref{1l}) possesses integrable dispersive regularisation which can be obtained by setting $y=-x$ in equations (\ref{disN=2}):
\begin{equation}\label{1ld}
u_{t}+uu_{x}+\frac{\epsilon}{2}\frac{T+1}{T-1}h_{xx}- \frac{\epsilon}{2}v_{xx}=0, \qquad v_{t}+vv_{x}+\frac{\epsilon}{2}\frac{T+1}{T-1}h_{xx}- \frac{\epsilon}{2}u_{xx}=0,
\end{equation}
(we refer to \cite{BLL} for  similar integrable systems of ILW type). System (\ref{1ld})  possesses the Lax pair which can be obtained from Lax pair (\ref{LaxdisN=2}) by substituting $\psi\to \psi e^{k x}$ and then setting $y=-x$:
\begin{equation}\label{L1ld}
(\epsilon \partial_x+v)T\psi=\lambda (\epsilon \partial_x+u)\psi,  \qquad
\epsilon \psi_t=\frac{\epsilon^2}{2}\psi_{xx}+(w-\frac{\epsilon}{2}h_x)\psi,
\end{equation}
where $w=\frac{\epsilon}{2}\frac{T+1}{T-1}h_{x}$ and $\lambda=e^{-\epsilon k}$ is a spectral parameter. Truncating  the expansion
$$
\frac{\epsilon}{2}\frac{T+1}{T-1}=\frac{\epsilon}{2}\frac{e^{\epsilon \partial_x}+1}{e^{\epsilon \partial_x}-1}=\frac{\epsilon}{2}\coth \left(\frac{\epsilon}{2}\partial_x\right)=\partial_x^{-1}+\frac{\epsilon^2}{12}\partial_x-\frac{\epsilon^4}{720}\partial_x^3+\dots,
$$
one can obtain nearly integrable dispersive approximations to system  (\ref{1ld}). 

Analogous integrable dispersive regularisations can be constructed for  multilayer shear flows with coinciding vorticities within each layer (such flows can model several submerged parallel jets). As an example let us consider the two-layer case (setting again $\omega=g=1$):
\begin{equation*}
u^i_{t}+u^iu^i_{x}+h_x=0,\qquad v^i_{t}+v^iv^i_{x}+h_x=0, \qquad h=\sum (v^k-u^k),
\end{equation*}
$i=1, 2$. This system possesses integrable dispersive regularisation 
\begin{align*}
u^1_t+ u^1 u^1_x + \frac{\epsilon}{2}\frac{T + 1}{T-1} h_{xx} - \frac{\epsilon}{2}(v^1 + v^2+ u^2)_{xx}=0,\\
u^2_t + u^2 u^2_x + \frac{\epsilon}{2}\frac{T + 1}{T-1} h_{xx} - \frac{\epsilon}{2}(v^1 + v^2 - u^1)_{xx}=0,\\ 
v^1_t + v^1 v^1_x + \frac{\epsilon}{2}\frac{T + 1}{T-1} h_{xx} - \frac{\epsilon}{2}(u^1 + u^2 + v^2)_{xx}=0,\\
v^2_t + v^2 v^2_x + \frac{\epsilon}{2}\frac{T + 1}{T-1} h_{xx} - \frac{\epsilon}{2}(u^1 + u^2 - v^1)_{xx}=0,
\end{align*}
which results from the second-order dispersive deformation constructed in Section \ref{sec:2n} by setting $y=-x$. The corresponding Lax pair is
\begin{equation*}
(\epsilon \partial_x+v^1)(\epsilon \partial_x+v^2)T\psi=\lambda (\epsilon \partial_x+u^1)(\epsilon \partial_x+u^2)\psi,  \qquad
\epsilon \psi_t=\frac{\epsilon^2}{2}\psi_{xx}+(w-\frac{\epsilon}{2}h_x)\psi,
\end{equation*}
where $w=\frac{\epsilon}{2}\frac{T+1}{T-1}h_{x}$ and $\lambda$ is a spectral parameter.

The general $n$-layer case possesses integrable dispersive regularisation
\begin{align*}
u^i_t+ u^i u^i_x + \frac{\epsilon}{2}\frac{T + 1}{T-1} h_{xx} - \frac{\epsilon}{2}(v^1 +\dots + v^n)_{xx}-\frac{\epsilon}{2}D^i_ju^j_{xx}=0,\\
v^i_t + v^i v^i_x + \frac{\epsilon}{2}\frac{T + 1}{T-1} h_{xx} -\frac{\epsilon}{2}(u^1 +\dots + u^n)_{xx}-\frac{\epsilon}{2}D^i_jv^j_{xx}=0,
\end{align*}
where $h=\sum (v^k-u^k)$ and $D^i_j$ is the $n\times n$ skew-symmetric matrix as in Section \ref{sec:2n}. The corresponding Lax pair is
\begin{equation*}
\prod (\epsilon \partial_x+v^k)T\psi=\lambda \prod (\epsilon \partial_x+u^k)\psi,  \qquad
\epsilon \psi_t=\frac{\epsilon^2}{2}\psi_{xx}+(w-\frac{\epsilon}{2}h_x)\psi,
\end{equation*}
where $w=\frac{\epsilon}{2}\frac{T+1}{T-1}h_{x}$ and $\lambda$ is a spectral parameter.

\section{Concluding remarks}

Here we list some problems for further study.
\begin{itemize}

\item It would be interesting to extend our classification to higher-order analogues of systems (\ref{ILW}).   
As an example let us consider the following $1$-parameter deformation of  the Lax pair (14):
\begin{align*}
T\psi&= \frac{\epsilon^2}{4}\psi_{yy} - \epsilon u \psi_y + \bigg(u^2 +{\epsilon}\lambda u_y  - 2v\bigg)\psi,\\
\epsilon \psi_t&= \frac{\epsilon^2}{4} \psi_{yy} + (2w - \epsilon u_y)\psi. 
\end{align*}
The compatibility conditions give the system
\begin{eqnarray}\label{TT}
u_t&=&uu_y+v_y+w_y-\epsilon \frac{2\lambda+1}{4} u_{yy},\\ \nonumber
v_t&=&2vu_y+uv_y+\epsilon \frac{2\lambda+1}{4}(v_{yy}+w_{yy}-u_y^2)-\epsilon^2 \frac{2\lambda^2+2\lambda+1}{8} u_{yyy},\nonumber
\end{eqnarray}
with the nonlocality $ \triangle w=\frac{T+1}{2}u_y$. Note that the third-order derivative $u_{yyy}$ disappears only for the particular parameter values  $\lambda=\frac{-1\pm i}{2}$, which leads to system (\ref{ILW5}). Under the differential substitution
$\tilde v=v-\epsilon \frac{2\lambda +1}{4}u_{y}$, equation (\ref{TT}) takes $\lambda$-independent form,
\begin{eqnarray*}
u_t&=&uu_y+\tilde v_y+w_y,\\ \nonumber
\tilde v_t&=&2\tilde vu_y+u\tilde v_y-\frac{\epsilon^2}{16} u_{yyy}.\nonumber
\end{eqnarray*}
Thus, there exist nontrivial higher-order  integrable systems with nonlocality of ILW type.
It would  be of interest to classify third-order integrable extensions of systems (\ref{ILW}).

\item Looking for integrable dispersive deformations of various dispersionless integrable systems, we came across the following (unforeseen) phenomenon: there exists (numerous) examples of 3D dispersionless integrable systems $\Sigma$ possessing integrable dispersive deformations  which, however, are not compatible with reductions  of $\Sigma$. As an illustration let us consider the dispersionless integrable system from Section \ref{sec:cont},
$$
u^i_{t}=u^iu^i_y+w_y, \quad w_{x}=(u^1+u^2+u^3)_{y},
$$
$i=1, 2, 3$, which possesses integrable dispersive deformation
\begin{align*}
u^1_t &= u^1u^1_y + w_y + \frac{\epsilon}{2}(u^2 + u^3)_{yy},\\
u^2_t &= u^2u^2_y + w_y + \frac{\epsilon}{2}(-u^1+u^3)_{yy},\\
u^3_t &= u^3u^3_y + w_y + \frac{\epsilon}{2}(-u^1 - u^2)_{yy},\\
\triangle w &= \frac{T + 1}{2}(u^1 + u^2 + u^3)_y.
\end{align*}
At the dispersionless level, this system possesses a reduction $u^3=u^2$,
\begin{equation}\label{def23}
u^1_{t}=u^1u^1_y+w_y,\quad u^2_{t}=u^2u^2_y+w_y, \quad w_{x}=(u^1+2u^2)_{y},
\end{equation}
which, however, is not compatible with the dispersive deformation. Attempts to seek integrable dispersive deformations of system (\ref{def23}) in the form 
\begin{align*}
u^1_t &= u^1u^1_y + w_y +\epsilon(\dots)+\epsilon^2(\dots)+\dots,\\
u^2_t &= u^2u^2_y + w_y +\epsilon(\dots)+\epsilon^2(\dots)+\dots,\\
\triangle w &= \frac{T + 1}{2}(u^1 + 2u^2)_y,
\end{align*}
where dots at $\epsilon^n$ denote differential polynomials of degree $n+1$ in the $y$-derivatives of $u^1, u^2$ and $w$, lead to a contradiction at the order $\epsilon^2$. 

It is a challenging problem to find out whether  the $2+1$ dimensional waterbag system (\ref{wb}) possesses  integrable dispersive deformations for rational values of the parameters $\omega_k$. This would provide a dispersive regularisation of  system (\ref{shear1}) governing  propagation of long nonlinear waves in a  shear flow with piecewise linear velocity profile (and rational vorticities).

\item Integrable dispersive regularisations of system (\ref{shear2}) constructed in this paper provide a framework which could be used  to  investigate the evolution of dispersive shock waves arising from suitable  initial conditions.

\end{itemize}

\section*{Acknowledgments}

We thank A. Chesnokov, S. Gavrilyuk, K. Khusnutdinova and A. Pogrebkov for useful discussions. 
The research of EVF was supported by a grant from the Russian Science Foundation No. 21-11-00006, https://rscf.ru/project/21-11-00006/.
The research of VSN was supported by the EPSRC grant EP/V050451/1.



\end{document}